# Bicarbonate/carbonic acid equilibria in aqueous solutions: Conformational Dynamics, Ion-Pair Speciation, and Free Energy Landscapes from Ab Initio Molecular Dynamics and Metadynamics

*Prabhat Prakash,[a] Karnamohit Ranka,[b] Frances A. Houle,[b] William A. Goddard III[a]*

a. Materials and Process Simulation Center, California Institute of Technology, Pasadena, California 91125, United States

b. Chemical Sciences Division, Lawrence Berkeley National Laboratory, Berkeley, California 94720, United States.

**ABSTRACT**

Carbonic acid ($H_2CO_3$) is connected with the dissolved $CO_2$ and bicarbonate ($HCO_3^-$) equilibria in aqueous solutions. Its two elementary reactions in water, acid dissociation and dehydration to $CO_2$, are difficult to resolve because both are coupled to proton transfer through hydrogen-bonded networks. In this work, we combine ≈ 0.5 ns of unbiased PBE-D3 molecular dynamics of 1 molar solution of $H_2CO_3$ with metadynamics on the O – H coordination ($H_2CO_3 \rightleftharpoons HCO_3^- + H^+$) and the C–O distance of the leaving hydroxyl ($H_2CO_3 \rightleftharpoons CO_2 + H_2O$). These unbiased trajectories show frequent cis–cis ⇌ cis–trans interconversion through long-lived bicarbonate (well separated from solvated proton, $H_3O^+$, and also mediated by a short-lived three-water [$H_2O\cdots H_3O^+\cdots H_2O$] ring that acts as a solvent-shared ion pair (SSIP); only 16-23% of deprotonation events reach a solvent-separated ion pair (SShIP). Resolving ion-pairs as SSIP for proper reaction leads to an improved estimate of $pK_{a1}$ from 1.1 to 2.4, toward the experimental 3.45-3.59 (measured at supposedly infinite dilution). Further, metadynamics simulations provide a dehydration barrier of 17.8 (1D) and 15.3 kcal/mol (2D), suggesting that the final product $CO_2 + H_2O$ is 2.8 kcal/mol below $H_2CO_3$. Our two-dimensional free energy surface also shows that C–O cleavage proceeds through the $CO_2 + OH^-$ which further goes down ~ 9.9 kcal/mol to product $CO_2 + H_2O$.

## INTRODUCTION

The $CO_2$/bicarbonate/carbonic acid equilibrium is one of the most ubiquitous chemical systems in nature as it governs processes ranging from ocean acidification and climate regulation to biological pH homeostasis and lab-scale to industrial carbon capture. When carbon dioxide dissolves in water, it establishes a series of coupled equilibria:

$$CO_2(aq.) + H_2O \rightleftharpoons H_2CO_3 \rightleftharpoons HCO_3^- + H^+ \rightleftharpoons CO_3^{2-} + 2H^+ \quad (1)$$

These seemingly simple equilibria exhibit remarkable complexity because speciation, proton transfer kinetics, and environmental factors such as pH, temperature, and ionic strength are coupled.[1,2] The carbonate system is the primary buffer of seawater, where equilibria among $CO_2(aq)$, $H_2CO_3$, $HCO_3^-$, and $CO_3^{2-}$ hold the surface pH near 8.1-8.3,[3] and of blood, where carbonic anhydrase interconverts $CO_2$ and $HCO_3^-$ to keep the physiological pH within 7.35-7.45.[4] In electrochemical $CO_2$ reduction ($eCO_2RR$), the same buffer plays a dual role: it provides a reservoir of $CO_2$-derived species and modulates the local pH at the electrode surface.[5–7] Thus, understanding the molecular mechanisms of $CO_2$ hydration, bicarbonate formation, and proton transfer is essential not only for fundamental chemistry but also for carbon capture, $CO_2$ conversion, and sustainable energy systems.[8–10]

While bulk-phase carbonate chemistry is relatively, well characterized experimentally, its behavior at catalytic interfaces, particularly electrode surfaces, goes beyond conventional modeling approaches. For example, during $eCO_2RR$, significant pH gradients develop within nanometer-to-micrometer scale diffusion layers adjacent to the electrode surface.[11,12] These gradients arise from the consumption or generation of protons during multiple proton-coupled-electron transfer (PCET) reactions. A key difficulty for molecular-scale modeling lies in the disparity of relevant length and time scales. Chemical reactions at active sites occur on nanometer and femtosecond-to-picosecond scales, while pH gradients extend over micrometers and develop over milliseconds to seconds. The electric double layer (EDL), which governs ion distribution and potential drop near the electrode, spans nanometers and influences both reaction thermodynamics and kinetics.[13–15] Further, the $CO_2$ concentration, pH, ion distribution, and water structure at catalytic interfaces deviate substantially from bulk conditions owing to the interplay between kinetics and mass transport.[16] Continuum and kinetic models that bridge these scales take the various rate and equilibrium constants of eq. 1 as input. As a result, an accurate molecular picture of the bulk equilibria, including species such as ion-pairs that do not appear in eq. 1 explicitly, is a prerequisite for predicting carbon speciation and reactivity at interfaces.

Though $H_2CO_3$ is kinetically stable, and water molecules catalyze its decomposition[17], in solution, two equilibria can be easily confused. The true first acidity constant refers to $H_2CO_3 + H_2O \rightleftharpoons HCO_3^- + H_3O^+$; with $pK_{a1} = 3.45 \pm 0.15$ and $3.49 \pm 0.05$ from ultrafast infrared spectroscopy[18,19], and stopped-flow spectrophotometry provides 3.59 averaged over 6.6-42.8 °C.[20] The more easily observed acidity of this acid, pKa = 6.35 is an apparent constant for $CO_2(aq) + 2H_2O \rightleftharpoons HCO_3^- + H_3O^+$. The corresponding free energies are $\Delta G \approx +4.7$ kcal/mol for $H_2CO_3 \rightarrow HCO_3^- + H^+$ and $\Delta G \approx -3.8$ kcal/mol for $H_2CO_3 \rightarrow CO_2 + H_2O$ with liquid water as reference (−6.2 kcal/mol in the 1 M convention used by Wang et al.[20]), and the measured dehydration rate corresponds to a barrier of $\Delta G^{\ddagger} \approx 15.7$ kcal/mol. Intact $H_2CO_3$ is also a viable proton donor toward biological bases[21,22],which makes its acidity relevant beyond the carbonate buffer.

First-principles methods, particularly density functional theory (DFT)-based ab initio molecular dynamics (AIMD) and machine-learning interatomic potentials (MLIPs) trained on AIMD data, can resolve the molecular mechanisms of $CO_2$ hydration and carbonate chemistry. Early AIMD studies explored: a water-wire-assisted hydration of $CO_2$ with a nucleophilic attack of $OH^-$ ion,[23,24] the hydrogen-bond structure of the carbonate species,[25] and the formation of $H_2CO_3$ via $HCO_3^-$ in water[26]. Finite size cluster calculations showed that additional water molecules lower the hydration and decomposition barriers.[27] Galib and Hanna[28,29] used Car-Parrinello MD with metadynamics on coordination-number collective variables (CVs) with 45 $H_2O$ and obtained $pK_{a1}$ = 2.6-3.8 for the ct and tt conformers; they also showed that $HCO_3^-$ decomposes to $CO_2 + OH^-$ only after full ion separation, with barriers of 24 - 27 kcal/mol, and that the dissociation barrier drops at the air-water interface. Recent AIMD-trained MLIP simulations with metadynamics revealed an "in-and-out" mechanism for $CO_2$ hydration at the air-water interface, where $CO_2$ must diffuse into the aqueous surface layer before reacting, and gave $pK_{a1}$ = 3.9 in bulk water.[30] Multi-level embedding approaches that combine correlated wavefunction theory with DFT-MD have shown that hydration energetics require dynamical solvent sampling.[31,32] Zhao et al.[33] showed that while the cis–cis (cc) conformer is the most stable, the minority cis–trans (ct) conformer is the dominant proton donor owing to its hydrogen-bonding network and electronic asymmetry.

Two aspects of such simulations complicate a direct comparison with concentrated systems. First, in a periodic cubic cell (used in AIMD) that is slightly above 1 nm size, the proton released by $H_2CO_3$ cannot move far from the bicarbonate, so that a substantial fraction of the "dissociated" state is a contact or solvent-shared ion pair, whereas the experimental $pK_{a1}$ refers to free ions at infinite dilution. Second, C-O cleavage and proton transfer are coupled, and a one-dimensional reaction coordinate can hide the slow proton degree of freedom.

In this work, we use PBE-D3 molecular dynamics (MD) and metadynamics simulations of $H_2CO_3$ in water, guided by three questions:

(i) What is the conformational and ion-pair landscape of carbonic acid and bicarbonate in water when sampled without bias?
(ii) What are the free energy profiles and molecular mechanisms of $H_2CO_3$ dehydration to $CO_2 + H_2O$, and which proton channel accompanies C--O scission?
(iii) How does the treatment of contact and solvent-separated ion pairs (CIP and SSIP) change the $pK_{a1}$ obtained from simulation?

To address the first question, we performed unbiased DFT-MD simulations of 1 M $H_2CO_3$ (for ≈0.5 ns in total) that capture the conformational flexibility of carbonate species in explicit water. Unlike static geometry optimizations or constrained sampling approaches, these unbiased MD trajectories reveal the dynamic interplay between intramolecular geometry changes and solvent reorganization to identify metastable ion pairs that serve as intermediates for conformer interconversion. For the second and third questions, we employ metadynamics on the O–H coordination of the carbonate oxygens and on the C–O distance of the leaving hydroxyl to reconstruct one- and two-dimensional free energy surfaces, resolve them by chemical species and ion-pair type, and quantify the activation barriers and reaction free energies that connect $CO_2$, carbonic acid, and bicarbonate.

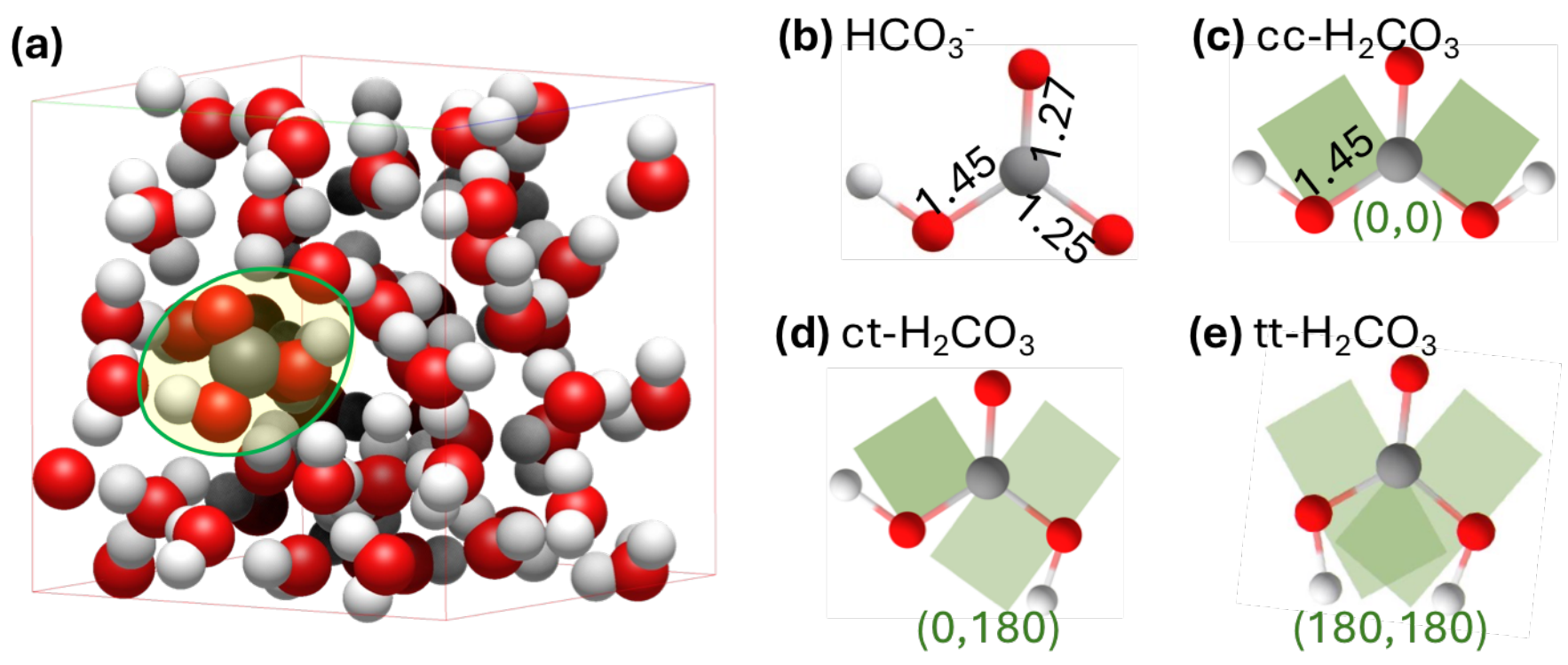

**Figure 1.** (a) Simulation cell with one cc-$H_2CO_3$ and 55 $H_2O$ (1 M). CCSD(T)-optimized gas-phase geometries of (b) bicarbonate anion $HCO_3^-$, (c) cis–cis (cc), (d) cis–trans (ct), and (e) trans–trans (tt) $H_2CO_3$. Green planes and labels show the H–O–C=O dihedral angles (degrees); numbers in black are bond lengths (Å). Atom colors: O, red; H, white; C, gray.

## COMPUTATIONAL DETAILS

### Unbiased DFT-MD

Two independent solvated systems, each containing one cc- or ct-$H_2CO_3$ and 55 water molecules in a cubic cell of 11.99 Å (0.99 M), were first equilibrated at 298 K for 500 ps with the all-atom OPLS[34] force-field for $H_2CO_3$ and the SPC/E model for water. All classical simulations were performed with LAMMPS (23 June 2022 version).[35] The equilibrated configurations were then simulated with DFT in the *NVT* ensemble using the Vienna Ab initio Simulation Package (VASP 6.4.3). The PBE functional[36] with the D3 dispersion correction[37] and projector-augmented wave (PAW) potentials were used with a plane-wave cutoff of 500 eV and Γ-point sampling. The temperature was maintained at 298 K with a Nosé-Hoover chain thermostat, and a time step of 1.0 fs was used for all MD and metadynamics simulations. The two trajectories, 276 ps (cc start) and 231 ps (ct start), sample repeated proton transfer between $H_2CO_3$ and water and the resulting interconversion of the cc and ct conformers via $HCO_3^-$.

Species were assigned frame by frame from the structure. A hydrogen atom was counted as bonded to a carbonate oxygen if their distance was below a cutoff (1.38 Å in the unbiased trajectories; 1.25

Å, the first minimum of the carbonate O-H distance distribution, in the metadynamics analysis) and that oxygen was the nearest oxygen of the hydrogen, so that a shared proton is never counted twice (Figure S1). Conformers of $H_2CO_3$ were assigned from the two H–O–C=O dihedral angles: (0°, 0°) for cc, (0°, 180°) for ct, and (180°, 180°) for tt (Figure 1c–e and Figure S2). $HCO_3^- \cdots H_3O^+$ ion-pairs were classified as contact (CIP), solvent-shared (SShIP), or solvent-separated (SSIP): by the number of water molecules bridging the hydronium and carbonate oxygens (0, 1, or ≥2), and by the hydronium-carbonate O···O distance $d_{ion}$ (CIP < 3.4 Å, SShIP 3.4–4.9 Å, SSIP ≥ 4.9 Å). A C–O bond was considered broken above 1.85 Å.

## Metadynamics

Free energy surfaces for acid dissociation and dehydration were obtained with metadynamics as implemented in VASP, with the electronic settings described above and an Andersen thermostat (with a collision probability of 0.10) at 298 K. Two CVs were used. CV1 is the total O–H coordination of the three carbonate oxygens ($k$), summed over all hydrogen atoms $i \in H_w \cup H_{ca}$ of water and carbonic acid:

$$\mathrm{CV1} = \sum_{k=1}^{3} \sum_{i \in \mathrm{H}} \frac{1-(r_{ki}/r_0)^{12}}{1-(r_{ki}/r_0)^{24}} \qquad (2)$$

where $r_{ki}$ is the distance between carbonate oxygen $k$ and hydrogen $i$, and $r_0$ = 1.38 Å. CV1 ≈ 2 for $H_2CO_3$ and ≈ 1 for $HCO_3^-$; because hydrogen-bonded protons contribute partially, CV1 is offset from the integer proton count (Figure S1). CV2 is the distance between the carbonate carbon and the oxygen that leaves on dehydration, $d$(C–O); its $H_2CO_3$ minimum lies at ≈1.40 Å and the $CO_2$ + $H_2O$ minimum at ≈3.0 Å.

One-dimensional (1D) runs were performed along CV1 (600 ps) for acid dissociation and along CV2 (1.2 ns) for dehydration, and a two-dimensional (2D) run biased CV1 and CV2 together (1.04 ns). Gaussian hills of 5 meV height and 0.05 (CV units) width were added every 50 fs, and repulsive walls (width 0.3; height 3 eV for CV1 and 5 eV for CV2) kept the CVs within the range of reaction sampling. To maintain the stochasticity of the thermostat, a new random seed was used after the addition of every 10 hills. The hydronium–bicarbonate distance $d_{ion}$ was not biased; its distribution was obtained by reweighting the CV1 trajectory.

Because plain metadynamics keeps adding bias, the instantaneous estimate $F(s,t) = -V(s,t)$ fluctuates by several hill heights. We therefore also report the bias averaged over the second half of each run, $F(s) = -\langle V(s,t) \rangle$, and use its drift over the last half of the run as the uncertainty. 2D surfaces were projected onto one CV by Boltzmann integration over the other. Barriers and reaction free energies were taken between the minima in the windows CV2 ≤ 1.7 Å (bound) and CV2 ≥ 2.3 Å (dissociated), and the $H_2CO_3$ and $HCO_3^-$ basins were split at CV1 = 1.5. Distributions in unbiased coordinates (species, $d_{ion}$) were obtained by weighting each frame with

$$w_t = \exp[-F(s_t)/kT]/N(s_t), \quad (3)$$

where $N(s)$ is the number of frames in the CV bin. Tiwary and Parrinello's weighting method[38] was used for comparison. Species-resolved free energies follow from $F_i(s) = F(s) - kT \ln P(i|s)$, where $P(i|s)$ is the fraction of species $i$ at a given CV value.

The pKa1 was then determined from the populations $P$ of the two species in the simulation cell of volume $V$:

$$\mathrm{p}K_{a1} = -\log_{10}(P_{\mathrm{HCO3-}}/P_{\mathrm{H2CO3}}) + \log_{10}(c^{\circ} N_A V) \quad (4)$$

Further details are given in Section S1 of the Supporting Information.

**Stochastic simulations**

We use stochastic chemical kinetics[39], to examine the time evolution of reactions using the rate constants derived from metaMD simulations. This model uses iterative and randomly selected events weighted by their probability. The Kinetiscope software package[40] was used to run these simulations. The selection frequency for events was adjusted between 10,000 – 1,000,000 depending upon the rarity of events. Further, for processes exceeding the diffusion limits, scaling factors were used to bring the process occurrence below diffusion limit, while maintaining the same thermodynamic ratio of entities.

## RESULTS AND DISCUSSION

**Conformer interconversion and ion pairs from unbiased MD**

The $CO_2/HCO_3^-$ equilibria in water involve reactions with low to high barriers, and only the low-barrier ones are accessible within a few hundred picoseconds of unbiased MD. To identify them, we simulated two 1 M systems starting from the cc and ct isomers of $H_2CO_3$ (Figure 1a) for 276 and 231 ps, respectively. In neither trajectory did the C–O bond break to form $CO_2$, consistent with a dehydration barrier above 15 kcal/mol (see below) despite the negative reaction free energy.

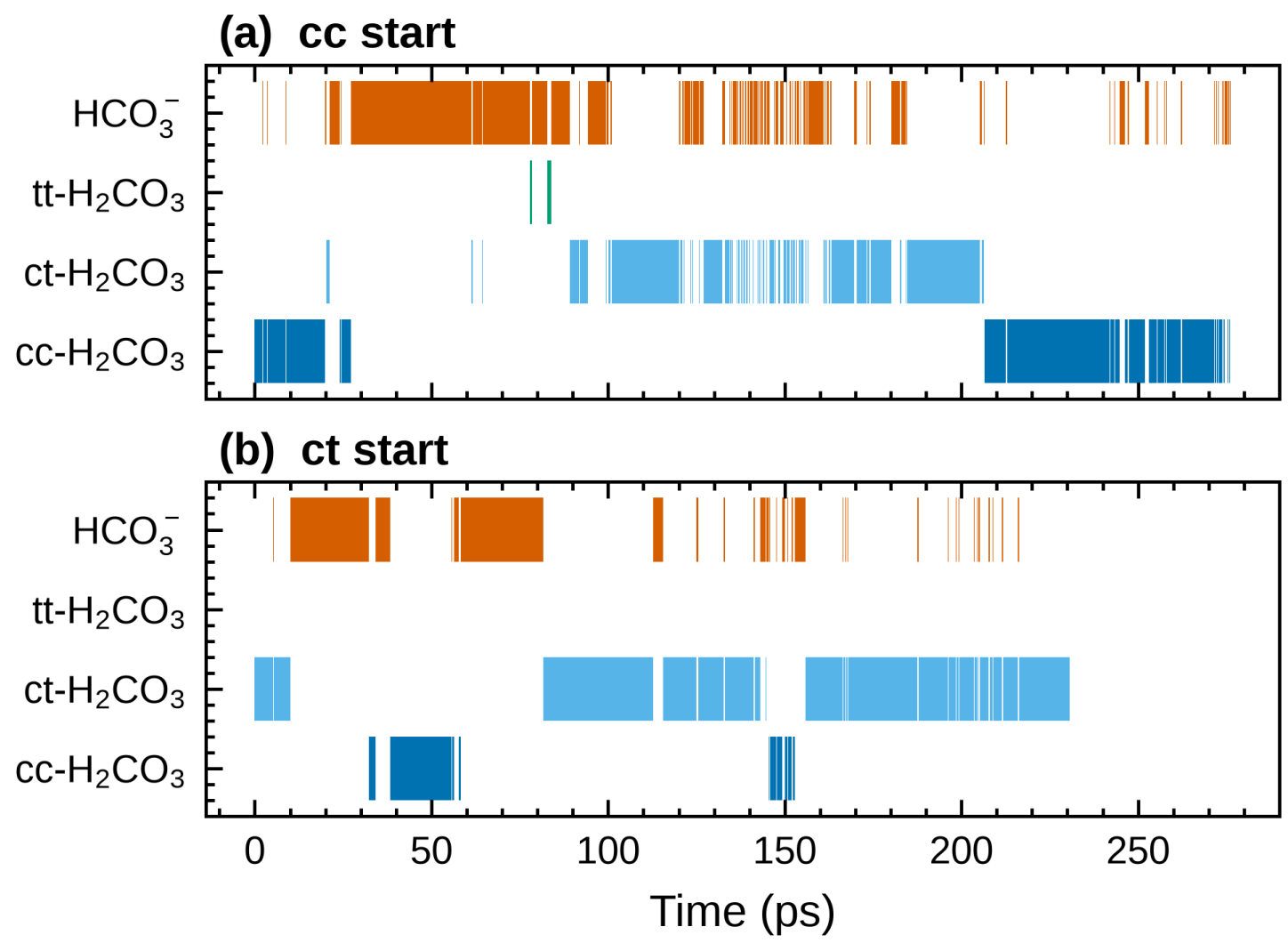

**Figure 2.** Species along the unbiased DFT-MD trajectories started from (a) cc- and (b) ct-$H_2CO_3$. Species were assigned from the number of O–H bonds on the carbonate oxygens and the two H–O–C=O dihedral angles; a 50 fs window was used to removes vibrational fluctuations.

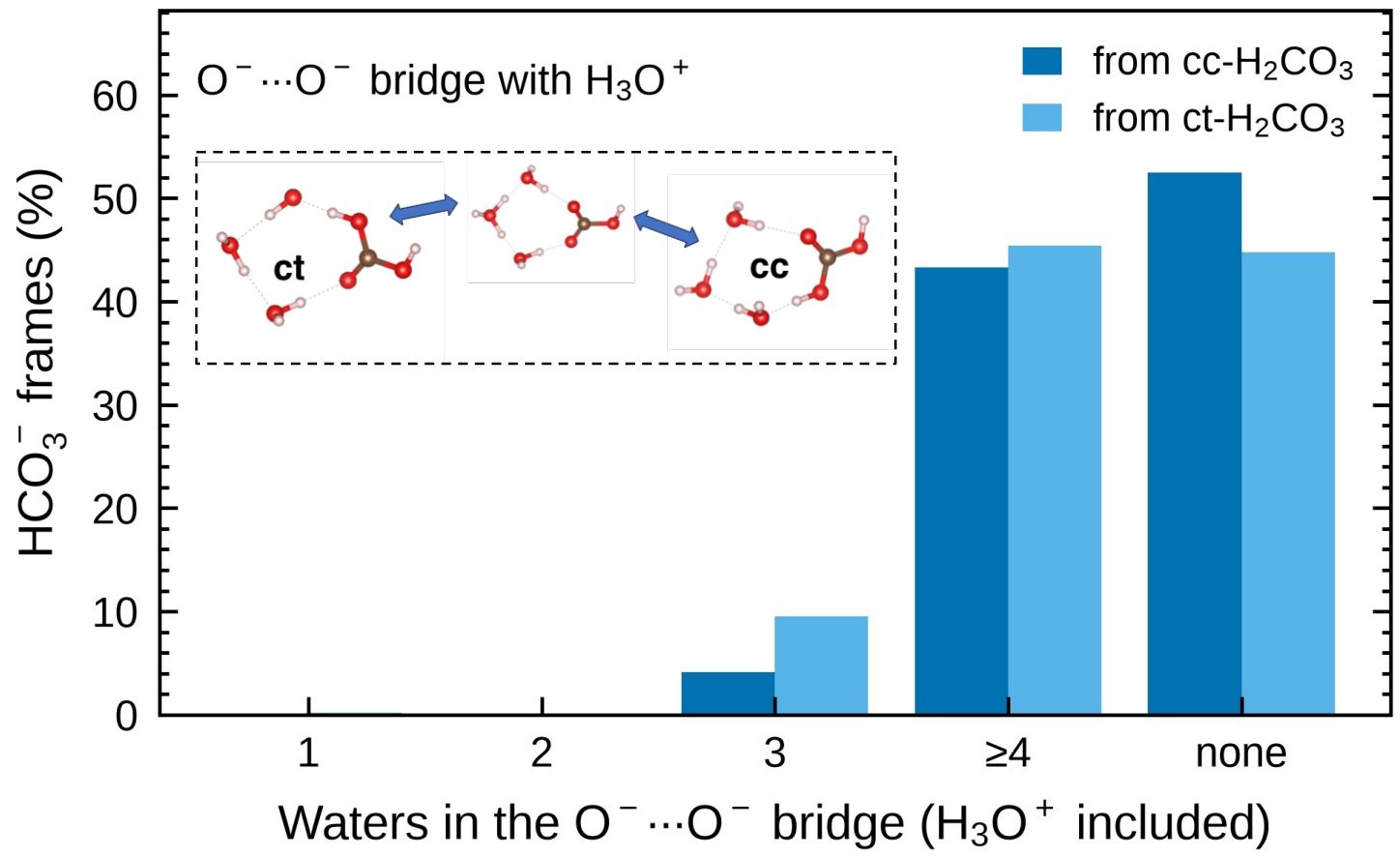

**Figure 3.** Number of molecules in the hydrogen-bonded bridge that connects two carbonate oxygens through $H_3O^+$ in the $HCO_3^-$ frames of the unbiased trajectories. The inset shows ct ⇌ cc interconversion through the three-water $[H_2O\cdots H_3O^+\cdots H_2O][HCO_3^-]$ ring, a solvent-shared ion pair (SShIP) in which the bicarbonate binds the bridge in a bidentate fashion.

In both trajectories, all species except the tt conformer appear in significant proportions and interconvert frequently (Figure 2 and Table S1). Starting from cc, the trajectory spends 30.8% of the time as cc, 29.9% as ct, and 38.8% as $HCO_3^-$; starting from ct, the corresponding values are 11.3%, 59.8%, and 28.9%, and tt appears in less than 1% of the frames. The interconversions occur in all permutations, i.e., cc ⇌ $HCO_3^-$, ct ⇌ $HCO_3^-$, and cc ⇌ ct via $HCO_3^-$. During cc ⇌ ct interconversion, $HCO_3^-$ is intermittently stable while the released proton diffuses into the solvent, and closed hydrogen-bonded rings form between two carbonate oxygens through $H_3O^+$ and water (Figure 3). Bridges of four or more molecules are the most common (43-45% of $HCO_3^-$ formed), whereas the three-water $[H_2O\cdots H_3O^+\cdots H_2O][HCO_3^-]$ ring is less frequent (4-10%) but is the geometry through which cc ⇌ ct interconversion most often proceeds (Figure 3, inset); rings larger than four molecules cannot be counted reliably because their size approaches the periodic boundary limits. The average lifetime of the three-water ring is 1.6 ± 1.1 ps, compared with 0.4 ± 0.3 ps for the four-membered bridge. We attribute the role of the three-water ring to two factors: (i) its longer lifetime gives the O–H bonds enough time to exchange their covalent and hydrogen-bonds, and (ii) the ring is a solvent-shared ion pair (SShIP) that serves as a metastable precursor for both cc and ct. This bridge is visited frequently in solution. In contrast, in a finite cluster with two additional water molecules, the $[(H_2O)_2\cdots H_3O^+\cdots(H_2O)_2][HCO_3^-]$ bridge requires $\Delta G^‡ = 11$ kcal/mol from ct to the transition state with gas-phase DFT **(Figure S4)**, which suggests that the surrounding solvent stabilizes SShIP. Concerted proton sharing through water rings and wires of this kind have been described for the hydrated excess proton and for water autoionization[41–43].

Because the SShIP mediates conformer interconversion, we analyzed the populations of all ion pairs. This distinction is also central to how $pK_{a1}$ should be computed, because contact ion pairs (CIPs) are easily miscounted as released protons, particularly in concentrated solutions. Of all the $HCO_3^-$ frames, 47% (cc start) and 59% (ct start) are SSIP, 23–25% are SShIP, and 16–30% are CIP (Table S1). However, only a minority of deprotonation events lead to separated ions: of the 102 and 51 deprotonation events in the two trajectories, 23 and 8 (16–23%) reach the SSIP state, and the remaining protons return to the bicarbonate from a contact or solvent-shared pair (Figure 4a,b). Figure 4c shows $pK_{a1}$ derived from these populations as a function of the minimum number of water molecules that separate $H_3O^+$ and $HCO_3^-$. Counting every dissociated configuration gives $pK_{a1}$ = 0.7 (cc start) and 1.0 (ct start); requiring at least one, two, or three intermediate water molecules raises the value to 1.1, 1.5, and 2.0, respectively, toward the experimental value of 3.45-3.59. The absolute values remain low, partly because PBE over-delocalizes the excess proton and over-stabilizes charge-separated configurations[44], and partly because ≈0.5 ns of unbiased sampling cannot converge to produce a good estimate of equilibrium constant. The trend, however, identifies the ion-pair state as a quantity that has to be controlled in any simulation estimate of $pK_{a1}$.

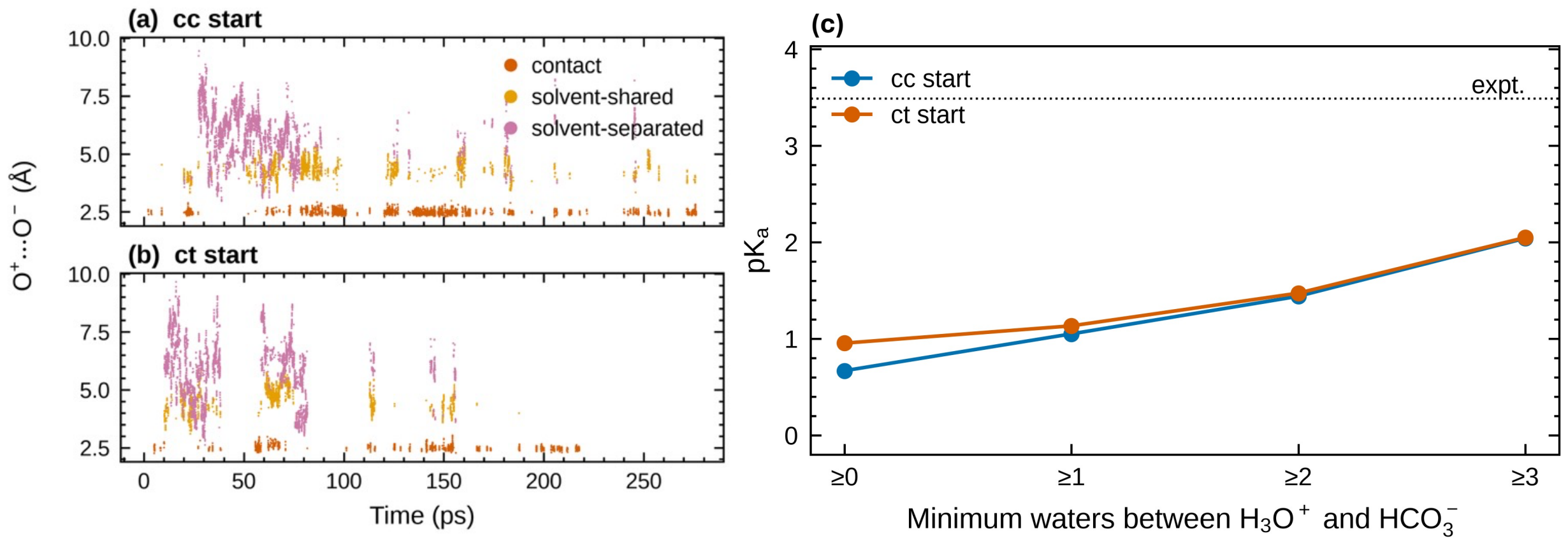

**Figure 4.** $H_3O^+\cdots HCO_3^-$ O···O distance along the unbiased trajectories started from (a) cc and (b) ct, colored by contact, solvent-shared, and solvent-separated ion pairs. (c) $pK_{a1}$ from the populations of the two trajectories as a function of the minimum number of water molecules between $H_3O^+$ and $HCO_3^-$; dotted line: experiment.

## $pK_{a1}$ from metadynamics

To obtain $pK_{a1}$, we ran metaMD simulations with CV1 for 600 ps and resolved the free energy by reweighting it for SSIP species **(Figure 5)**. In this case, when every ion pair (including CIP and SShIP) is counted, $\Delta G$ ($HCO_3^- + H_3O^+ \rightleftharpoons H_2CO_3$) is close to zero **(Figure 5a)**. The $d_{ion}$ criteria used along the progress of CV1 **(Figure S5)** shows that the $HCO_3^-$ side of the reaction is populated almost entirely by CIP and SShIP, consistent with the unbiased trajectories, in which most of the deprotonation events do not form SSIP. However, the TP reweighting with SSIP gives a free energy barrier of 0.97 kcal/mol for the forward reaction and 3.33 kcal/mol for the backward reaction. Resolving ion pairs therefore shifts the computed $pK_{a1}$ by more than two units (0.01 to 2.37), towards the experimental value. It is important to note here, that the reaction free energy (2.36 kcal/mol) would only produce a pKa1 = 1.73, however, the value 2.37 is derived using the population in both sides of the reaction using eq.4. The representative structural snapshots for all important conformers, transition states and IPs are provided in **Figure S6a,b of the SI.**

The derived values of barriers also carry a substantial sampling uncertainty. **Table S2** shows both free energy based and TP reweighting based estimates and range of fluctuation in barriers and pKa1. The more important observation is that the difference between the "any" and "well-separated" ion-pair definitions lead to a difference of 2.3 pKa units. This difference is larger than just the PBE functional error and originates in the finite cell rather than in the electronic structure.

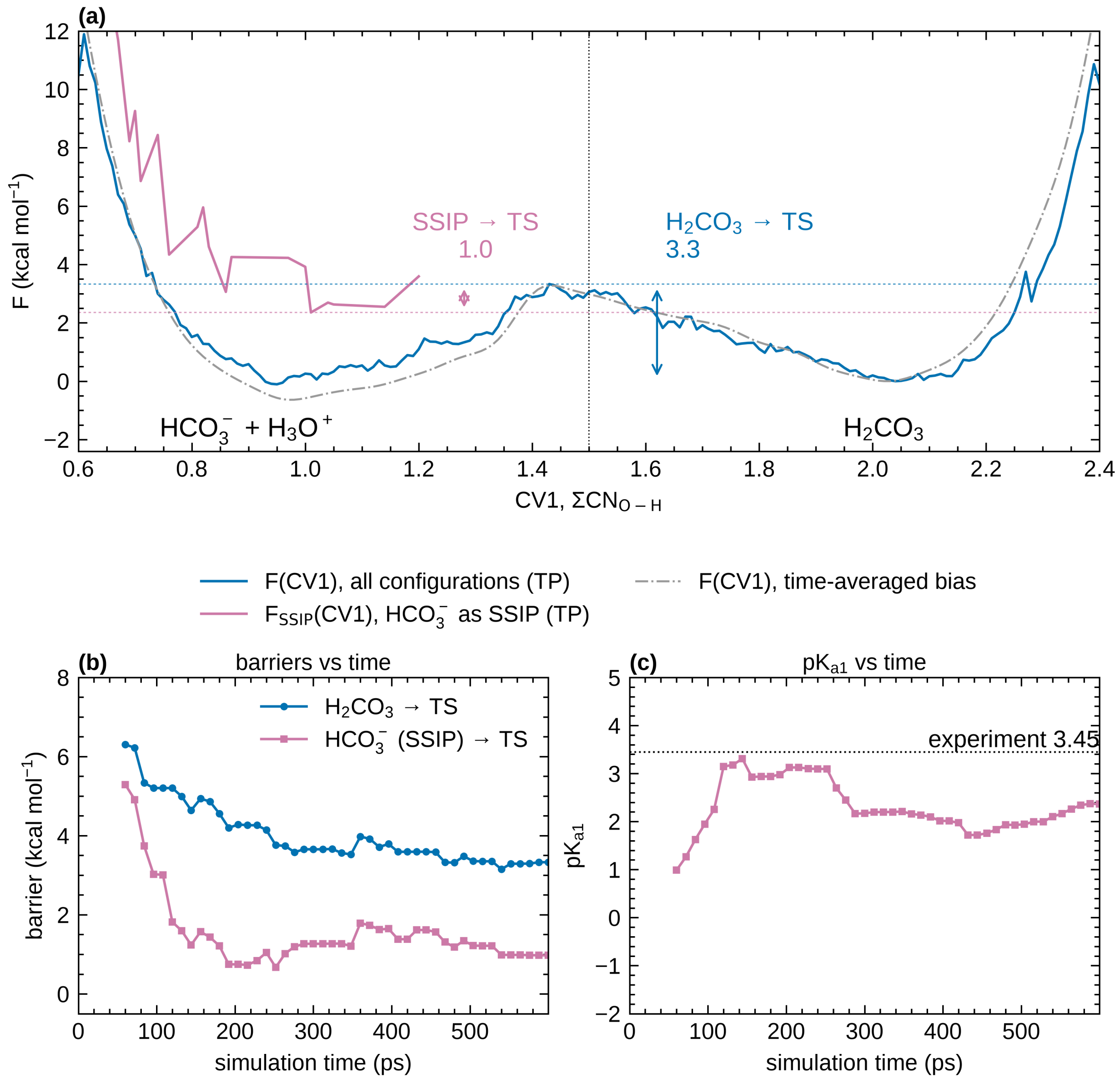

**Figure 5.** Species-resolved free energy of $H_2CO_3$ dissociation from the 600 ps CV1 run: (a) Reweighted free energy along CV1 over all configurations (blue) and for the SSIP only (purple), time-averaged bias (grey). (b) Forward and backward free-energy barriers and (c) pKa1 progression with simulation time.

**Table 1.** $pK_{a1}$ of $H_2CO_3$ ($H_2CO_3 + H_2O \rightleftharpoons HCO_3^- + H_3O^+$, ≈298 K) from experiment and simulation. "Any": every proton transfer to water counted; "separated": solvent-separated ions with the accessible-volume correction. An extended comparison is given in Table S4.

| Source | Method | $pK_{a1}$ |
|---|---|---|
| Adamczyk et al.(a) | ultrafast IR | 3.45 ± 0.15 |
| Pines et al. (b) | ultrafast IR/fluorescence | 3.49 ± 0.05 |
| Wang et al. (c) | stopped-flow (6.6–42.8 °C) | 3.59 |
| Galib and Hanna (d) | CPMD metaMD, BLYP, 45 $H_2O$ | 2.6–3.8 (ct, tt) |
| Brookes et al. (e) | revPBE-D3 MLIP, multiple walkers metaMD | 3.9 |
| this work | unbiased PBE-D3 MD, populations | 2.0 |
| this work | PBE-D3 metaMD (1D), 600 ps | 2.37 |

a. Ref. 18 ; b. Ref. 19 ; c. Ref. 20 ; d. Ref. 28 ; e. Ref. [30]

## Dehydration of $H_2CO_3$ to $CO_2 + H_2O$

Unlike the deprotonation discussed in the previous section, dehydration reaction $H_2CO_3 \rightarrow CO_2 + H_2O$ is not accessible to unbiased MD. To quantify it, we biased the C–O distance of the leaving hydroxyl (CV2) for 1.2 ns. This run escaped the $H_2CO_3$ well for the first time after 40.9 ps and crossed between the bound and dissociated windows 26 times; after the first return at 218 ps, the median interval between crossings is 30 ps **(Figure S6)**. The time-averaged profile **(Figure 6a)** gives a dehydration barrier of $\Delta G^\ddagger$ = 17.8 kcal/mol with the transition state at $d$(C–O) = 1.92 Å, and places $CO_2 + H_2O$ at 3.01 Å, 2.8 kcal/mol below $H_2CO_3$; the reverse (hydration) barrier is 20.6 kcal/mol. The final bias alone gives 18.4 and −6.8 kcal/mol, and the difference between the two estimates is a measure of the remaining uncertainty of plain metaMD. Experiment gives $\Delta G^\ddagger \approx$ 15.7 kcal/mol and $\Delta G \approx -3.8$ kcal/mol, which imply a hydration barrier of ≈19.5 kcal/mol **(Table 2)**.

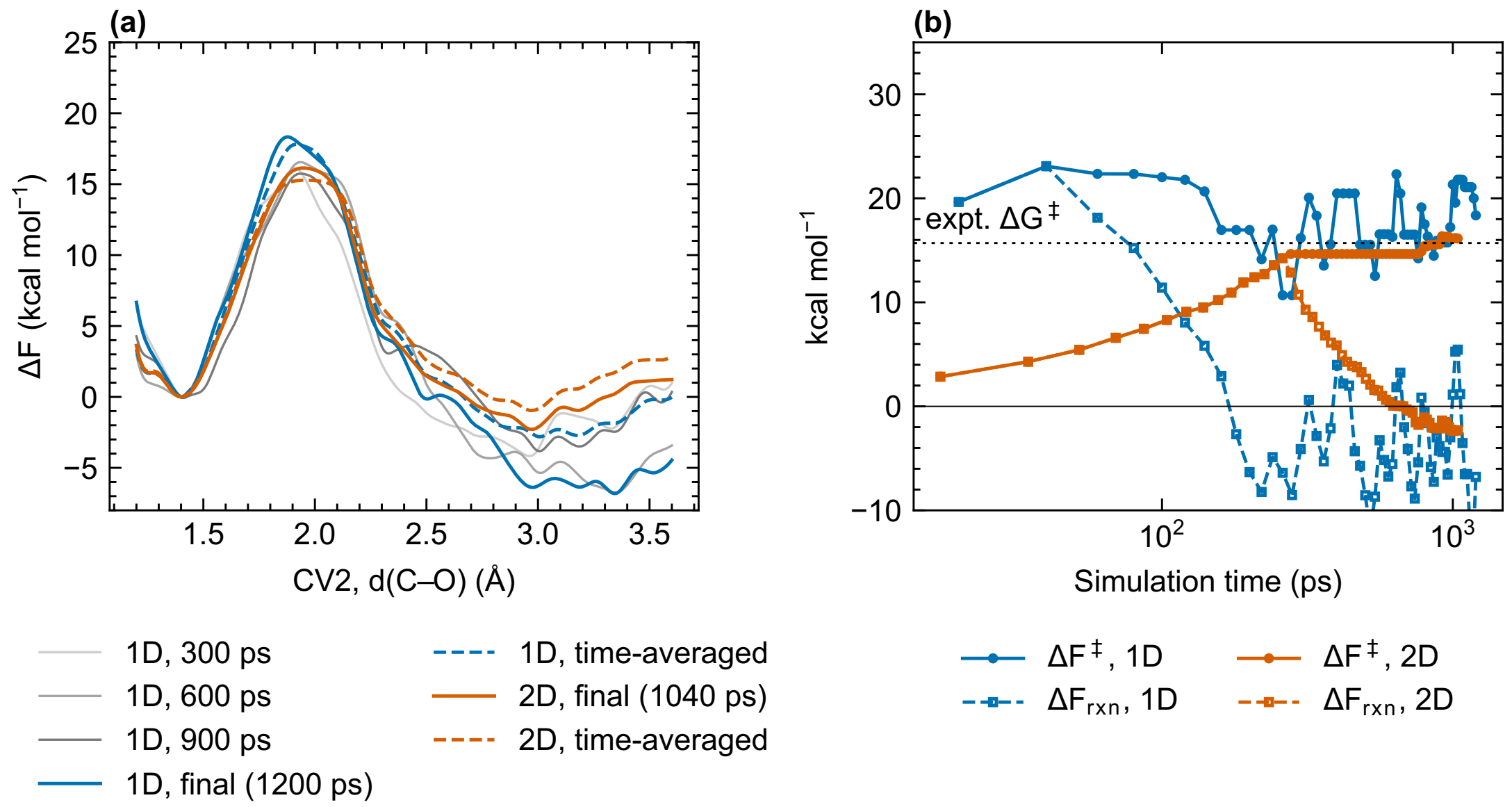

**Figure 6.** Free energy along the C–O distance. (a) 1D run biased on CV2: profiles after 300, 600, and 900 ps, final (1200 ps), and time-averaged. (b) 2D run projected on CV2. (c) Barrier $\Delta F^{\ddagger}$ and reaction free energy $\Delta F_{rxn}$ against simulation time for the 1D and 2D runs; dotted line: experimental dehydration barrier.

However, the 1D profile does not reveal which entity, $H_2CO_3$ or $HCO_3^-$ accompanies C–O scission. Thus, a 2D surface $F$(CV2, CV1) from the 1.04 ns run **(Figure 7a)** separates the four species. Relative to $H_2CO_3$, the basin minima are {$HCO_3^-$} at +2.0, {$CO_2$ + $H_2O$} at −1.2, and {$CO_2$ + $OH^-$} at +9.9 kcal/mol (representative snapshots for each of these shown in **Figure 7b-e**). The scission of the C-O bond therefore proceeds such that, the leaving OH group takes up a proton after the C–O bond breaks. The anionic $CO_2$ + $OH^-$ part shows the reverse of $OH^-$ attack on $CO_2$ which lies 10 kcal/mol uphill.[23,32] This agrees with Galib and Hanna,[45] who found that $HCO_3^-$ decomposes to $CO_2$ + $OH^-$ only after full ion separation and with a barrier of 24–27 kcal/mol, and with the water-assisted decomposition proposed from finite sized cluster calculations.[27] By projecting the 2D metaMD on CV2, the 2D run gives $\Delta F^{\ddagger}$ = 15.3 kcal/mol and $\Delta F$ = −1.0 kcal/mol **(Figure 6b)** also implying that the $CO_2$ + $H_2O$ well is probably not fully filled after 1.04 ns because the trajectory should ideally also sample the proton coordinate more. All the free energy values from both 1D and 2D runs are provided in **Table 2**.

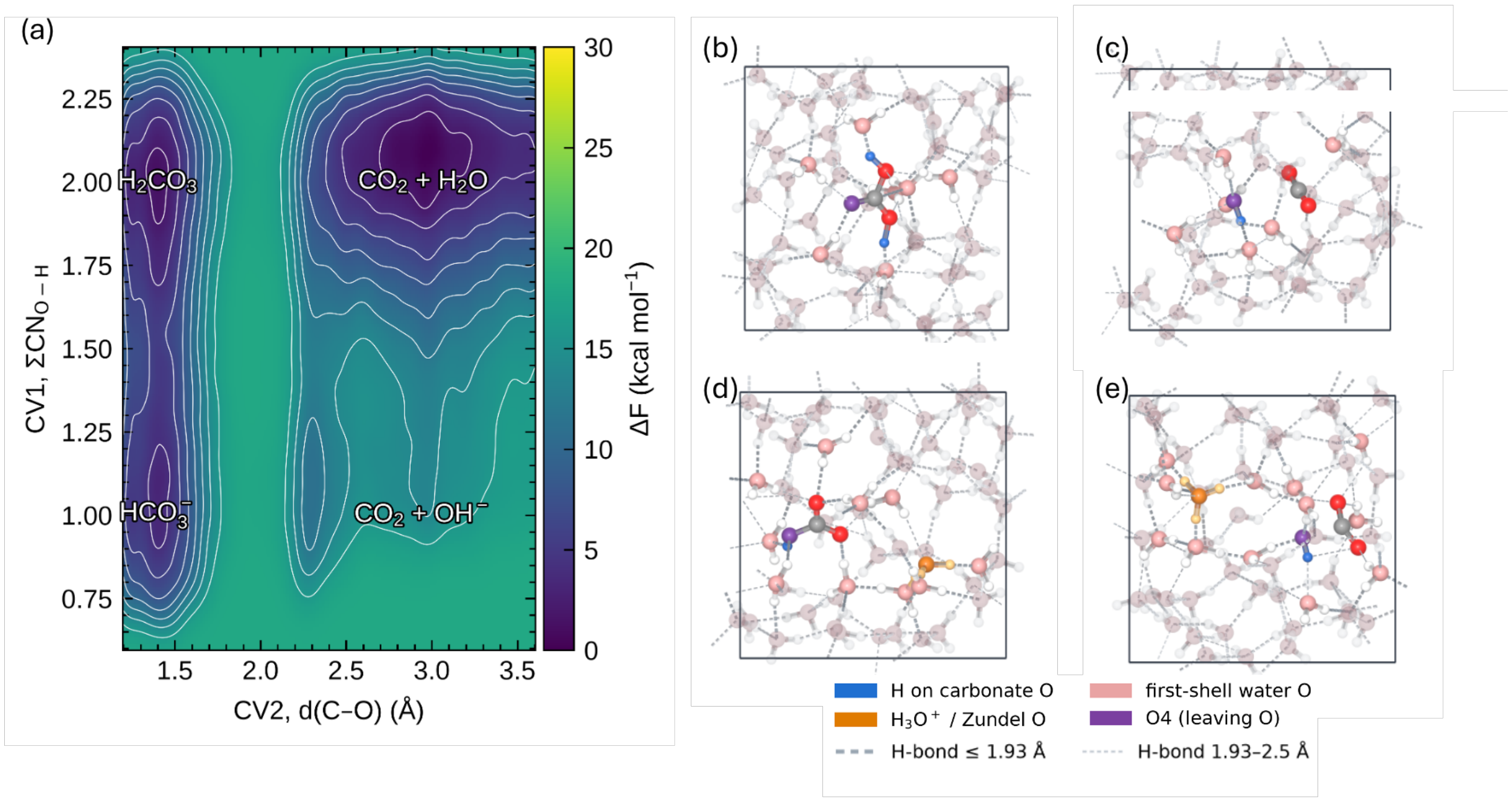

**Figure 7.** (a) Two-dimensional free energy surface $F$(CV2, CV1) of $H_2CO_3$ dehydration from the 2D run (time-averaged bias); contours every 2 kcal/mol. CV1 is the O–H coordination of the carbonate oxygens, CV2 the C–O distance of the leaving oxygen. (b) – (d) are representative snapshots of the observed entities in four labeled basins showed in panel *a*.

**Table 2.** Free energies (kcal/mol) of $H_2CO_3$ dehydration from the time-averaged metaMD bias, referenced to $H_2CO_3$.

| Quantity | 1D (CV2) | 2D (CV1, CV2) | Experiment |
|---|---|---|---|
| $\Delta F^‡$, $H_2CO_3 \rightarrow CO_2 + H_2O$ | 17.8 | 15.3 | 15.7 ($\Delta G^‡$) |
| $\Delta F^‡$, $CO_2 + H_2O \rightarrow H_2CO_3$ | 20.6 | 16.3 | ≈19.5 |
| $\Delta F$, $CO_2 + H_2O$ (profile) | −2.8 | −1.0 | −3.8 |
| $CO_2 + H_2O$ (basin minimum) | — | −1.2 | −3.8 |
| $HCO_3^- + H_3O^+$ (basin minimum) | — | +2.0 | +4.7 (from $pK_{a1}$) |
| $CO_2 + OH^-$ (basin minimum) | — | +9.9 | — |

Resolving the 1D run by species **(Figure S7)** also shows that the reactant part is a mixture of $H_2CO_3$ and $HCO_3^-$ ($HCO_3^-$ +0.2 kcal/mol), that the dissociated side is $CO_2 + H_2O$ (−3.7 kcal/mol), and that $CO_2 + OH^-$ appears only near the transition state (+9.1 kcal/mol). Accordingly, the 1D reaction free energy (−2.8 kcal/mol) is closer to experiment than the 2D estimate, whereas the 1D barrier (17.8 kcal/mol) lies above the 2D value (15.3 kcal/mol) because the $H_2CO_3$ basin was over-filled to 22.9 kcal/mol before the first escape, and plain metaMD removes such an excess of sampling only slowly. Over the last half of the 1D run, the barrier varies between 15.4 and 23.7 kcal/mol and $\Delta F$ between −12.0 and +6.8 kcal/mol **(Figure S6b)**, due to the uncertainty of a single plain-metaMD run with 26 crossings.

The two barriers from 1D and 2D runs are slightly above and below the experimental barrier 15.7 kcal/mol, respectively. Correlated-wavefunction embedding has shown that revPBE-D3 underestimates this barrier (0.53 vs 0.68 eV),[31], thus a PBE-D3 barrier clearly above experiment would point to residual sampling error rather than to the functional. Combining the two steps gives the apparent constant of the $CO_2$ system. With the 1D reaction free energy, $\log K_h = -2.1$ and $pK_{app} = pK_{a1} - \log K_h \approx 4.4$ for separated ions (2.1 counting all ion pairs), compared with 6.35 from experiment. Of the ≈1.9-unit deficit, about 1.1 units come from $pK_{a1}$ and 0.7 units from the error in $CO_2/H_2CO_3$ step (−2.8 vs −3.8 kcal/mol).

### Kinetic consequences for the carbonate system

We define four reactions as listed in Table S2. We used three models for comparison, experimentally derived rate constants, Brookes et al.'s most recent numbers from significantly long simulations with MLIP, and DFT-metaMD derived numbers from this work. In our dataset, for the third reaction, $k_r = 2.26 \times 10^{10}$ $s^{-1}$, and $k_f = 5.24 \times 10^{12}$ $M^{-1}s^{-1}$. Since $k_f$ is significantly above the diffusion limit, in the stochastic simulation model, we divide both these numbers by 10,000 to keep below diffusion limit while keeping the effective $pK_{a1}$ unchanged.

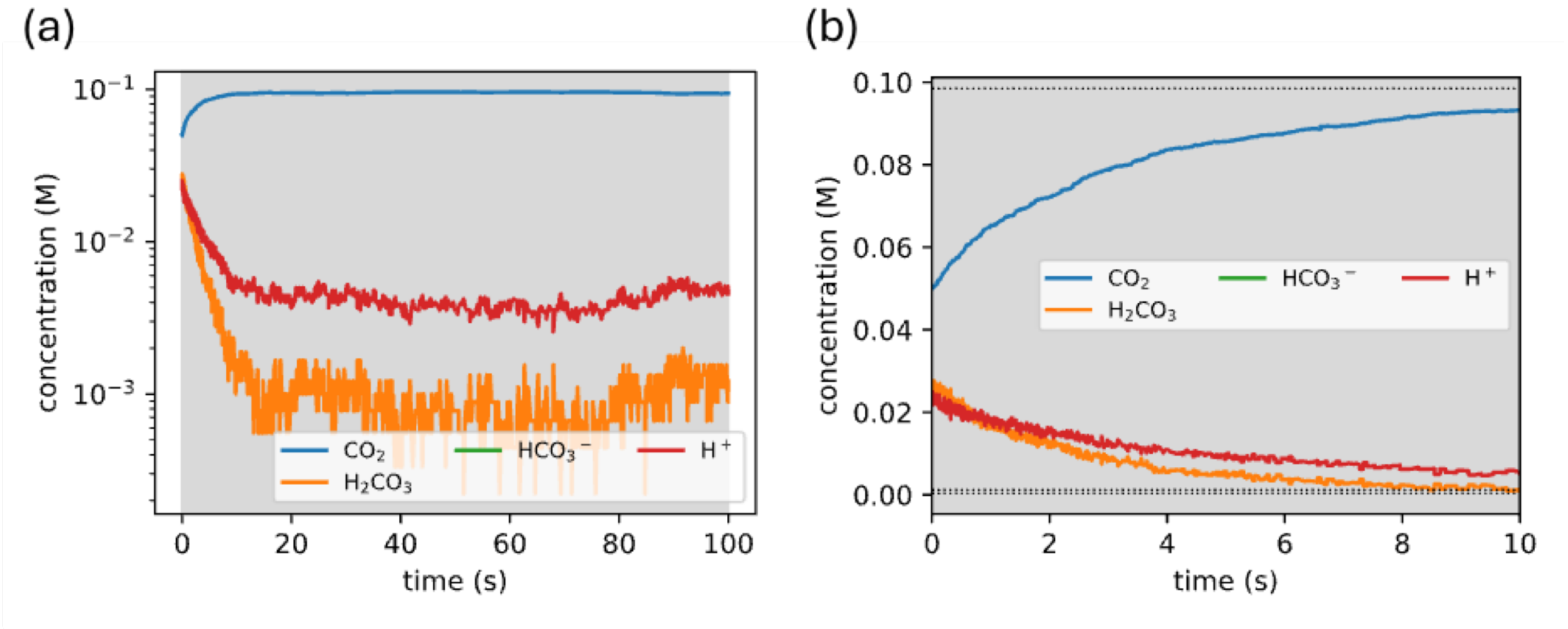

**Figure 8.** Evolution of concentration *c* of entities started with initial $CO_2$ at 0.05 M, $HCO_3^-$, $H_2CO_3$ and $H^+$ each at 0.025 M, equivalent to total carbon concentration of 0.1 M, with **(a)** log *c* vs. full simulation time, **(b)** *c* vs. first 10 second time, using the rate constants derived from CV1 and CV2 (1D) metaMD simulations, from stochastic simulations. The concentration of $HCO_3^-$ and $H^+$ ions is exactly same, thus overlapped in both plots. Reaction schemes and rate constants are provided in **Table S3**. A comparison with other stochastic simulations from experimental, another MD simulation work, and 2D metaMD derived rate constants is provided in **Figure S9**.

The scholastic simulation (ran for 100 seconds total time) shows that, the initial concentrations of all carbon species starts to saturate within first 10 seconds and completely saturate after 20 s **(Figure 8)**. For the 0.1 M system simulated here, with 1D metaMD on CV1 derived set, the predicted pH is 2.39 (using concentrations averaged from 80 – 100 s), compared to, 3.66 from experiment **(Figure S9)**, and 3.96 from other recent simulation work.[30]

We account for this disagreement of more than 1 pH unit to

i) GGA delocalization error as PBE overstabilizes delocalized, charge-separated states such as the $HCO_3^- \cdots H_3O^+$ pair and the Zundel ions, which underestimates pKa1;

ii) Further one $H_2CO_3$ in 55 waters (about 1 M) samples a contact or solvent-separated pair, not ions at infinite separation;

iii) The 1D and 2D barriers for step 4 differ by 4.3 kcal/mol (20.6 vs. 16.3), which indicates slow orthogonal degrees of freedom such as water-wire reorganization.

A more accurate functional and a bigger system size will be used in future to resolve these sources of errors.

## CONCLUSIONS

We used unbiased DFT-MD and metadynamics with PBE-D3 to resolve the dissociation and dehydration of aqueous $H_2CO_3$ at the level of individual ion pairs and protonated/hydroxide pathways. The unbiased trajectories show that cc- and ct-$H_2CO_3$ interconvert through $HCO_3^-$ on a picosecond time scale and that a three-water [$H_2O\cdots H_3O^+\cdots H_2O$] ring, a solvent-shared ion pair with a lifetime of 1.6 ps, mediates most of these events; only 16-23% of deprotonations reach a solvent-separated pair. For the dehydration reaction, metaMD with 26 recrossings gives a barrier of 17.8 (1D) and 15.3 kcal/mol (2D), bracketing the experimental value 15.7 kcal/mol, and suggests that {$CO_2$ + $H_2O$} is 2.8 kcal/mol below $H_2CO_3$ (experiment: 3.8 kcal/mol). The 2D metaMD also shows that the scission of C–O bond proceeds through immediate $OH^-$ recombining with proton, as $CO_2$ + $H_2O$, with {$CO_2$ + $OH^-$} +9.9 kcal/mol higher. Resolving ion pairs raises the computed $pK_{a1}$ from 0.1 to 2.4, roughly one unit below experiment, and the residual difference is consistent with the known over-stabilization of delocalized protons by PBE. The main limitation of the present estimates is the sampling uncertainty of single-walker plain metadynamics. Future work will use well-tempered, multiple-walker metadynamics and correlated-wavefunction corrections on configurations sampled and machine-learned from each basin.

## ASSOCIATED CONTENT

**Supporting Information**. Extended computational details (CV switching function, species and ion-pair assignment, accessible-volume correction, reweighting); species populations and dihedral distributions of the unbiased trajectories; cluster transition state; convergence of the metadynamics runs, reaction schemes and additional results from stochastic simulations.

## AUTHOR INFORMATION

### Corresponding Author

*E-mail: fahoule@lbl.gov , wag@caltech.edu

### Author Contributions

The manuscript was written through contributions of all authors. All authors have given approval to the final version of the manuscript.

### Funding Sources

Department of Energy, Office of Science, Office of Basic Energy Sciences, DE-SC0021266, DE-AC02-05CH11231, BES-ERCAP0024109

### Notes

The authors declare no competing financial interest.

## ACKNOWLEDGMENT

This work was supported by the Liquid Sunlight Alliance, which is supported by the US Department of Energy, Office of Science, Office of Basic Energy Sciences, Fuels from Sunlight

Hub under award no. DE-SC0021266. This research used resources of the National Energy Research Scientific Computing Center; a DOE Office of Science User Facility supported by the Office of Science of the U.S. Department of Energy under Contract No. DE-AC02-05CH11231 using NERSC award BES-ERCAP0024109 for some calculations. Some computations were conducted on the Resnick High-Performance Computing Center, a facility supported by Resnick Sustainability Institute at the California Institute of Technology.